\documentclass[fleqn]{article}
\usepackage{graphicx}
\usepackage[figuresright]{rotating}
\usepackage{axodraw}
\usepackage[T1]{fontenc}
\begin{document} 

%
{\centering \Large \bf
Software Tools for the Analysis of the \mbox{Photocathode}} \\

{\centering \Large \bf
Response of Photomultiplier Vacuum Tubes}

\vspace{1.0cm}  
{\centering \large 

Riccardo Fabbri$^{a}$\\

\vspace{0.3cm}      
{\large 
  $^a$ Forschungszentrum Juelich (FZJ), J\"{u}lich \\
  E-mail: r.fabbri@fz-juelich.de\\
}}

  \vspace{1.5cm}
  \hspace{3.7cm} 
{\centering \large 
  \today
}

\vspace{1.5cm}
\begin{abstract}
  The central institute of electronics (ZEA-2) in the 
  Forschungszentrum J\"{u}lich (FZJ) has developed a system 
  to scan the response of the photocathode of photomultiplier 
  tubes (PMT). The PMT sits tight on a supporting structure,  
  while a blue light emitting diode is moved along its surface 
  by two stepper motors, spanning both the $x$ and 
  $y$ coordinates. All the system is located in a light-tight 
  box made by wood. 

  A graphical software was developed in-situ to 
  perform the scan operations under different configurations (e.g., 
  the step size of the scan and the number of measurements per point). 
  During each point measurement the current output generated in 
  the vacuum photomultiplier is processed in sequence by 
  a pre-amplifier (mainly to convert the current signal into 
  a voltage signal), an amplifier, and by an ADC module
  (typically a CAEN N957). The information of the measurement is 
  saved in files at the end of the scan. 

  Recently, software based on the CERN ROOT~\cite{ROOT} and 
  on the Qt libraries~\cite{QT} was
  developed to help the user analyzing deeper the data obtained by 
  the scan. The new software, cross-platform due to its build performed
  inside the Qt-IDE, is described in this note. 

\end{abstract}

{\large \vspace{-18cm} \hspace{-4cm} 
  Forschungszentrum J\"{u}lich\\

  \hspace{-4.cm}
  Internal Report No. FZJ\_2013\_02988
}

\newpage  

\tableofcontents
  \vspace{-20cm} \hspace{6cm} 
\newpage  

\section{Introduction}
%
This note describes how to use the programs designed to analyze 
the result of the scan of a PMT photocathode (or of a series of scans), 
which I developed for the Neutron and Gamma Detector Group 
(Arbeit Gruppe Neutron und Gamma Detektoren) 
at \mbox{ZEA-2} in FZJ. 

\section{The Photocathode Scan Facility at ZEA-2}
\begin{figure}[b!]
      \includegraphics[height=9cm,width=13.cm]{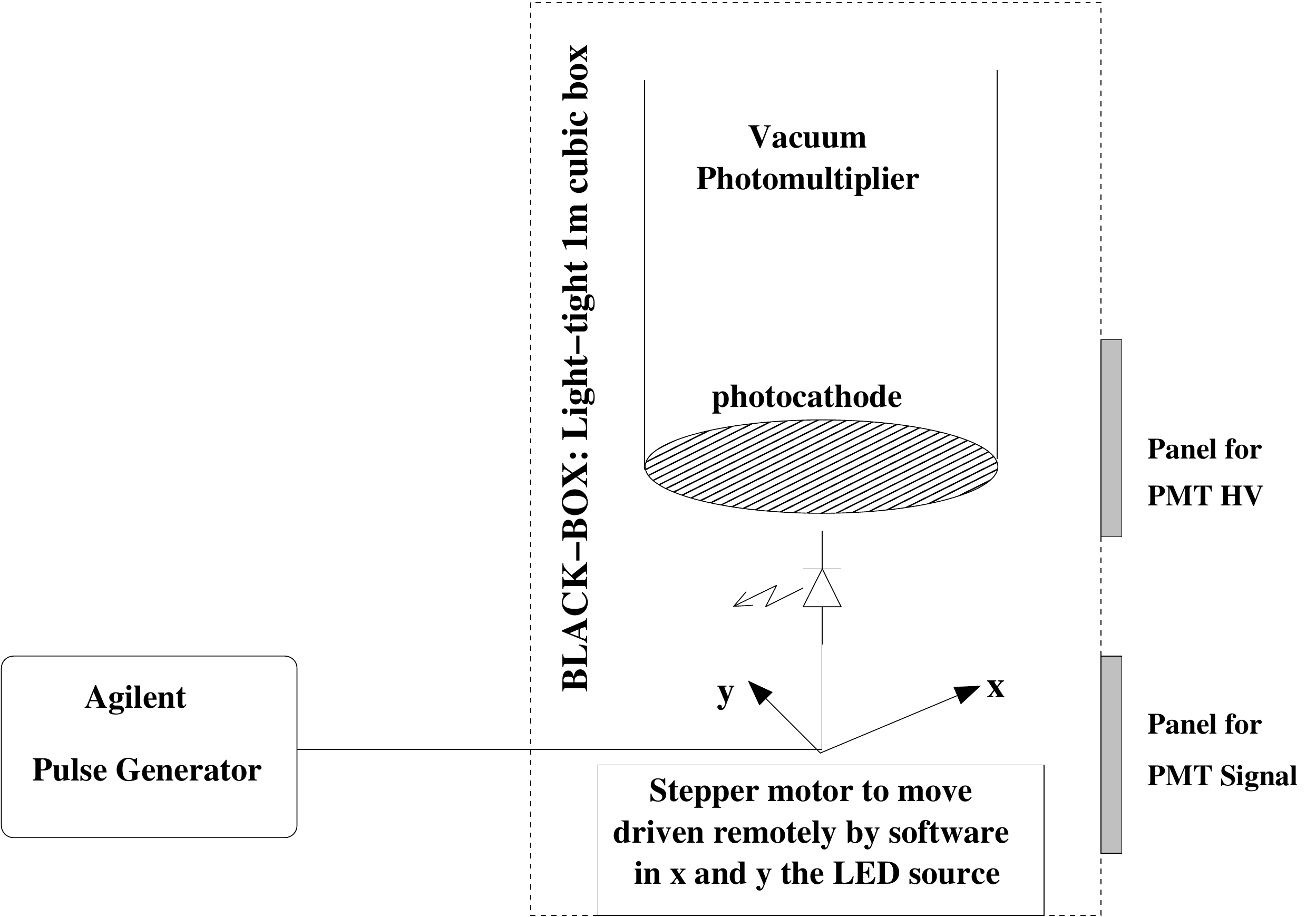}
  \caption{Schematic of the facility available to scan the response 
           of the photocathode surface of PMTs.}
  \label{fig:SCAN_BOX}
\end{figure}

%
In the institute of electronics ZEA-2 at FZJ since many years
a facility to analyze the performance of PMTs is present
and available to external users upon request. In principle,
photomultipliers of every size can be accommodated on the scan 
table inside the light-tight box, Fig.~\ref{fig:SCAN_BOX}.
The workshop of the institute can build the suitable support
to keep sitting fest any PMT photocathode in front of the
moving LED.

The scan is performed remotely by the program DetScan 
(designed and written by Uwe Clemens, ZEA-2 at FZJ), which drives the
stepper motors, and takes care 
to show either online or offline the intensity (in ADC units) 
of the measurement in the scanned area 
of the photocathode, and to save the measurement data in files
on disk. These files can be analyzed by the programs which I 
recently developed to investigate additional aspects of the 
photocathode response in a photomultiplier. 

The pulse generator is typically set to generate a voltage 
pulse such to measure enough signal in all the investigated 
area, without saturating the photomultiplier.  
The current signal from the PMT is converted into a voltage 
signal by a pre-amplifier located inside the box, and can be 
eventually readout via a front panel provided with LEMO connectors
with $50$ Ohm impedance.
The signal is additionally amplified and shaped by the 
ORTEC amplifier 571, and then sent to the CAEN ADC module N975.
The date acquisition can run in auto mode continuously 
sampling the incoming signals (above a user-defined low-level
threshold), and shipping the their peaking amplitude
to the DetScan program. Or, a gate, synchronized with the pulse to the LED,
can be set to the ADC module.
In this latter case the measurement of the pedestal (to be subtracted 
from the data) is needed, because in some outer regions of the photocathode 
surface the efficiency of the photo-electron emission could be low enough 
to have statistically no signal at the device output. 

It is clear that increasing the high voltage of the PMT can avoid this 
problem; on the other side, the region with more efficiency (also 
referred as gain, i.e., as the number of photo-electrons eventually
produced per 
impinging photon) might generate a large signal to saturate the ADC module.

\section{PMT Photocathode Scan:\\ Analysis of a single Scan Measurement}
\begin{figure}[b!]
  \hspace{-2cm}
      \includegraphics[trim=0cm 16.5cm 0cm 1cm, clip=true,
                       height=10cm, width=16.cm]
                       {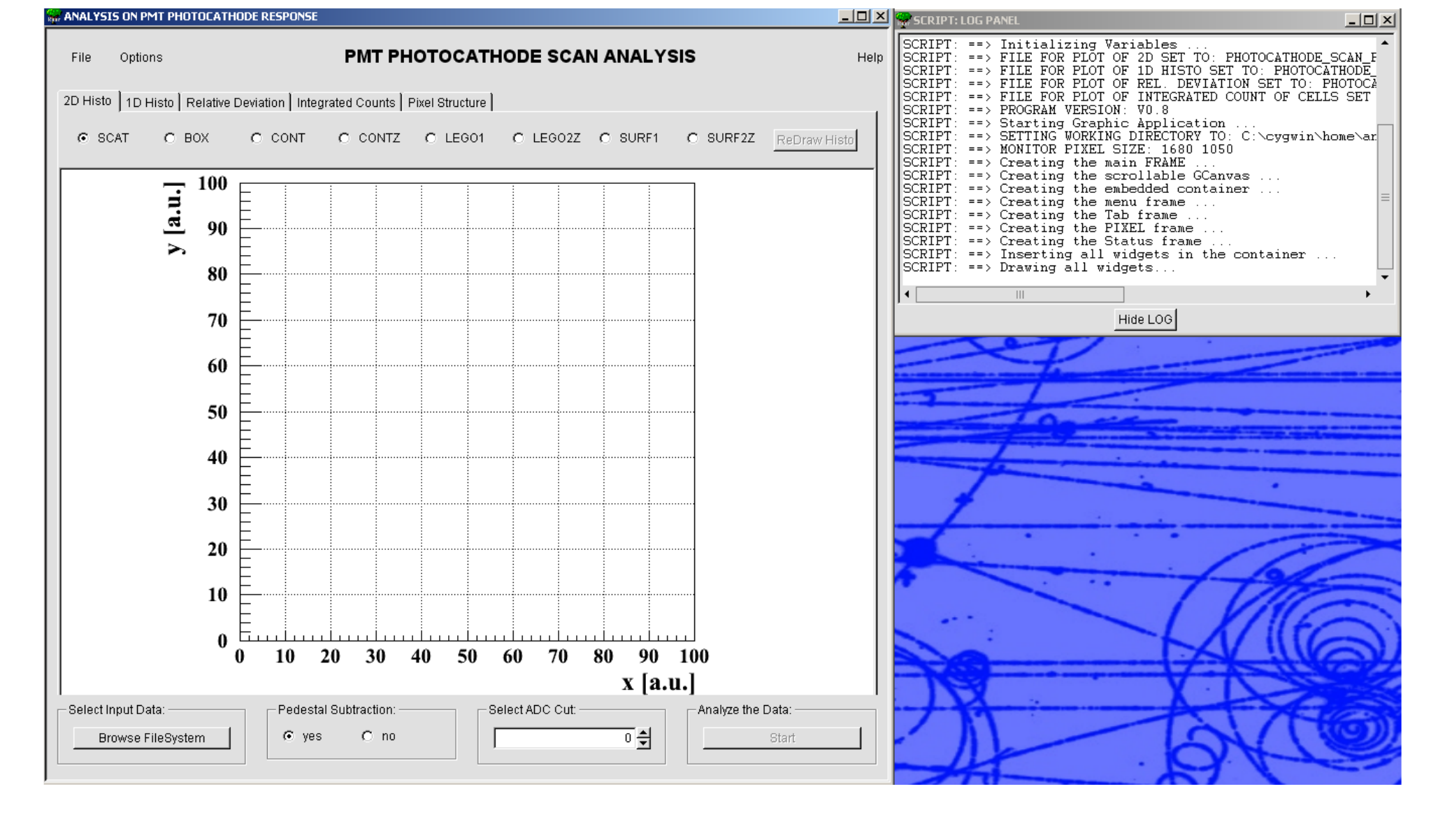}

  \vspace{-0.8cm}
  \caption{Screenshot at start-up of the GUI to analyze the 
           data provided by the 
           DetScan program. The program 
           is started along with LOG file canvas at its side.}
  \label{fig:PMT_SCAN_GUI}
\end{figure}

%
The PMT\_PhotoCathodeScan program
(written in C/C++ and based on the CERN ROOT libraries),
shown in Fig.~\ref{fig:PMT_SCAN_GUI},  
processes the data generated during the scan of a PMT 
photo-cathode by the DetScan program.

To start a measurement the user should select the input data 
via the button \underline{\it Browse FileSystem} in the bottom 
left side of the canvas;
the selection allows only to choose one of the *.csv files 
generated by the DetScan, which saves in these files 
(in a 128x128 mask) the mean amplitude measured in each 
scan step.
If an error in finding the input file is found, then that
error will be prompted in a pop-up window.  Instead, 
if no error appears (e.g. the selected file does exist) the 
button \underline{\it Start} will get activated and the data can 
be analyzed by pressing it.

Note that to access all the information relative to that
measurement all the corresponding files dumped 
by the DetScan program should be present in the same directory, 
as (among the other ones):\\

\hspace{-2cm}
\begin{tabular}{|l|l|}
    \hline
                             & Variable retrieved \\
    \hline
    $\star$\_Info.txt        & Start and stop time of the measurement \\
    $\star$\_TempIn.txt      & Temperature (Celsius degree) internal to the\\
                             & scan box measured during each scan step\\
    $\star$\_TempOut.txt     & Temperature (Celsius degree) in the laboratory\\
                             & measured during each scan step\\
    $\star$\_PedSpk.txt      & Histogram of the pedestal measurement \\
    $\star$\_<date:YYYYMMDD>\_<time:HHMM>.ph  & 
                             Timestamp of the measurement \\
                             & \\
    \hline
\end{tabular} \\ \\

After selecting the data file to process the user has the option
to whether subtract the pedestal calculated by the DetScan (in the
future it could be calculated directly from its saved spectrum) and 
to select a lower-level ADC cut. 
By clicking on the button \underline{\it Start} the analysis 
is performed 
with the selected options. In case a different set of options is
needed, then the analysis should be restarted.

The plots showed in the tabs can be saved in external files. This
possibility is  displayed in the menu option \underline{\it File}. The default 
graphic format is png but it can be changed via the menu option 
\underline{\it Options/Config Format for Pictures}.
In the same menu item \underline{\it Options} one can choose whether 
to show the legends in the plots, and to show the LOG file. 
At the end of the menu 
bar the option \underline{\it Help} gives the possibility to see the version 
history of the program and the reference to its author.

The program allows to show several data property in a 2D and 3D
visualization with different plotting options, as displayed in the 
radio-buttons
widgets in the top side of the tabs. In this case there is no need to 
restart the analysis, but it is sufficient to redraw the histogram with 
the button \underline{\it ReDraw Histo}. 
The available options are taken by the ROOT library. For clarification 
of the way all the ROOT objects used in this program behave please refer 
to the ROOT documentation~\cite{ROOT}.

\begin{figure}[ht!]
  \hspace{-2cm}
      \includegraphics[height=10cm, width=16.cm]
                       {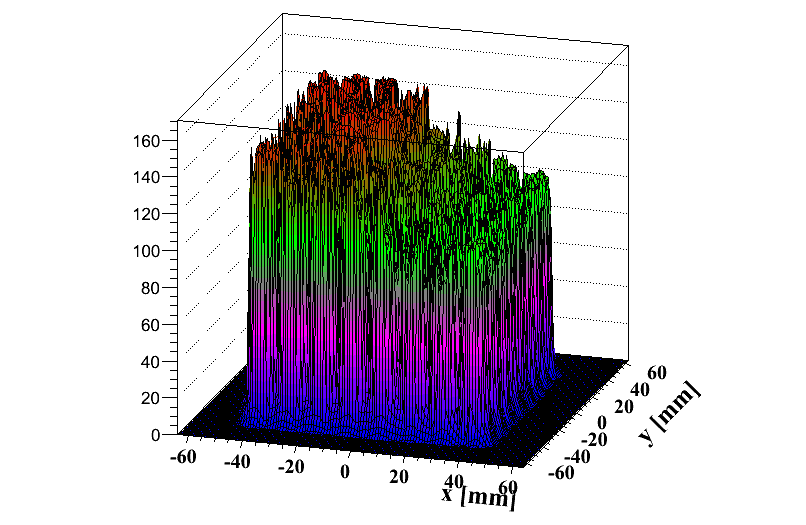}

  \vspace{-0.7cm}
  \caption{Intensity response of the photocathode surface for 
           the H8500 PMT.}
  \label{fig:PMT_SCAN_GUI_EXAMPLES}
\end{figure}

The Tab {\underline {\it 2D Histo}} shows the x-y gain response of the 
photo-cathode surface.
For every point scanned by the blue-light LED of the scanning
system the mean amplitude (as calculated by the DetScan) is shown 
(eventually with the pedestal subtraction in case this option was 
required). It is clear, that using the functions of the ROOT
classes, specific slides of the surface can be investigated
directly from the 2-dimensional histogram of the photocathode response.
An example of this measurement is shown in Fig.~\ref{fig:PMT_SCAN_GUI_EXAMPLES}
using the multi-anode 8x8 photomultiplier H8500 from 
Hamamatsu~\cite{HAMAMATSU}.
Clearly visible is the large non-homogeneous response on the scanned
surface. This is a typical behavior observed in PMTs.     

The Tab {\underline{\it 1D Histo}} shows the one-dimensional histogram 
which contains the amplitudes for all $x$ and $y$ bins.

The Tab {\underline {\it Relative Deviation}} shows the entries of the 
relative deviation  with respect to the measured maximum amplitude. 
Note that, by construction, the relative deviation can vary between 
null and unity.

\begin{figure}[ht!]
  \hspace{-1cm}
  \includegraphics[height=8cm, width=14.cm]
                   {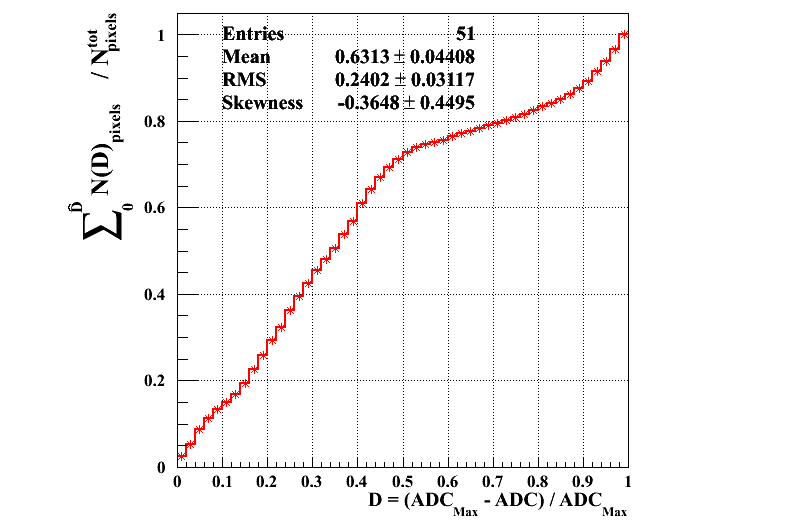}
  \caption{Integration of the relative deviation with respect to the 
           maximum amplitude measured in a scanned bin for 
           the H8500 PMT.}
  \label{fig:PMT_SCAN_GUI_INTEGRAL}
\end{figure}

The Tab {\underline {\it Integrated Counts}} shows the integrated sum of 
entries (pixels) with respect to the relative deviation. As an example, 
this histogram can tell the user what is the fraction of pixels within 
a relative deviation of value $D$. It is clear that considering that 
$D$ can range between $0$ and $1$, then the fractional integral can be 
at most unity, as shown in 
Fig.~\ref{fig:PMT_SCAN_GUI_INTEGRAL} obtained again using 
the H8500 photomultiplier. Please note that here the term pixel is 
interpreted not as a real physical pixel in the photocathode surface 
(as for example in multi-anode PMTs), but as the investigated region 
during one scan-step.
  
In this specific example, within a specific PMT and experimental setup, 
it appears that
$80\%$ of the scanned points have a response lower at least than 
$30\%$ of the measured maximum.

\begin{figure}[t!]
  \includegraphics[ height=8cm, width=12.cm]{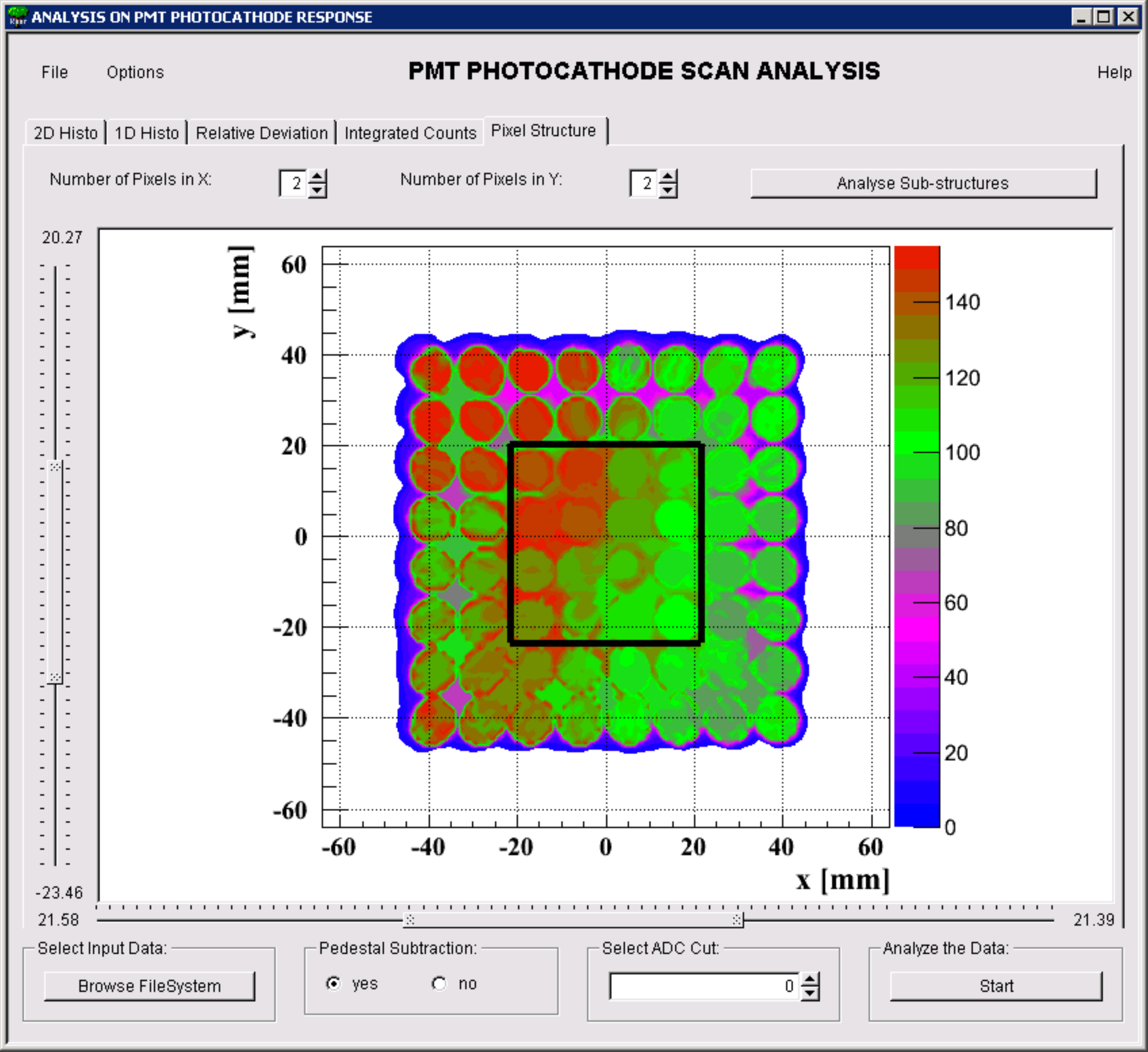}\\

  \vspace{-0.2cm}
  \includegraphics[ height=8cm, width=12.cm]{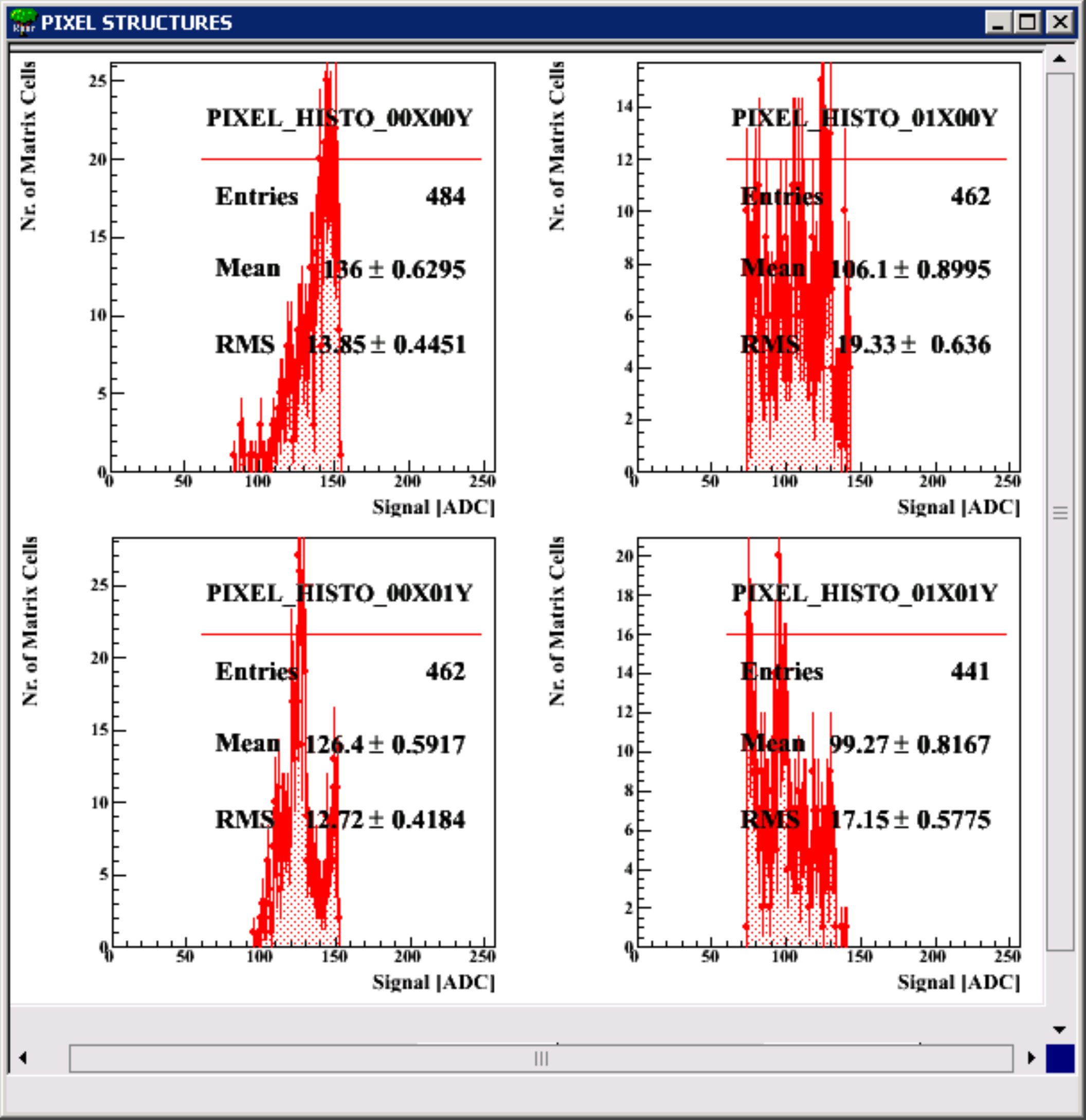}\\

  \vspace{-0.6cm}
  \caption{Top panel: The response of the H8500 multi-anode photomultiplier 
           with a plexi-glass mask on its surface 
           (which results in the observed circular 
           shapes) shows the strong non-homogeneous response
           of the device. Bottom panel: The scrollbars select
           the area to be divided in several pixels whose
           intensity is shown in the corresponding histogram.}
  \label{fig:pixels}
\end{figure}

The Tab {\underline {\it Pixel Structure}}, 
upper panel of Fig.~\ref{fig:pixels}, is very useful to 
compare the response of different area (of equal size) of the 
photocathode surface. 

The scrollbars in $x$ and $y$
allow to select the area of interest, which can be then divided in sub-pads 
using the corresponding {\underline {\it Number of Pixels in X (Y)}}
entry widgets. 
When the selection is done, then clicking on the button 
"{\it Analyse Sub-structures}" pops up a new canvas 
displaying all the selected pads; each pad shows a one-dimensional 
histogram containing the amplitude value of all scanned points in the 
corresponding region, bottom panel of Fig.~\ref{fig:pixels}. 

The variation of the mean value of 
the displayed histograms shows the gain mismatch in different area of the 
photocathode surface. In case of a multi-anode PMT, as the H8500, 
this tool can quantify within few percent of systematical uncertainty
the gain mismatch between the independent amplifying
channels (properly selecting the area under 
investigation, and the number of equal-size pixels in $x$ and $y$),
thus allowing for its correction by introducing, e.g., adequate
resistors in the channels with higher gain.

Moving downward the scrollbar the two buttons {\it Hide Canvas} and
{\it Dump Results} at the bottom of the canvas become visible.
They allow, respectively, to iconize the frame, and to dump the 
results into an ascii file whose path is dumped in the LOG panel. 
The result file can be eventually used for an offline analysis
on the gain-matching. To perform this analysis it is crucial
to correctly select the area to investigate by setting the 
area borders as close as possible to the transition regions 
between the physical pixels.

\section{PMT Scan Systematics:\\ Stability of the Experimental Apparatus}
%
The stability of the scan measurement is a crucial 
aspect to monitor, in order to reliably provide the 
results of a photocathode scan. On this regard,
a GUI was developed to analyze a set of measurements 
performed in the same hardware configuration. This
allows to monitor how stable is the measurement (e.g.,
the mean intensity of the PMT), and in case of any 
observed instability, to possibly spot a dependence
of the measurement on some variables of the system.
Please note that the measurements to compare should be
performed while keeping the system as much as possible 
untouched, to avoid introducing a bias in the 
measurement.

\begin{figure}[t!]
   \hspace{-3cm}
      \includegraphics[height=12cm,width=18.cm]
                      {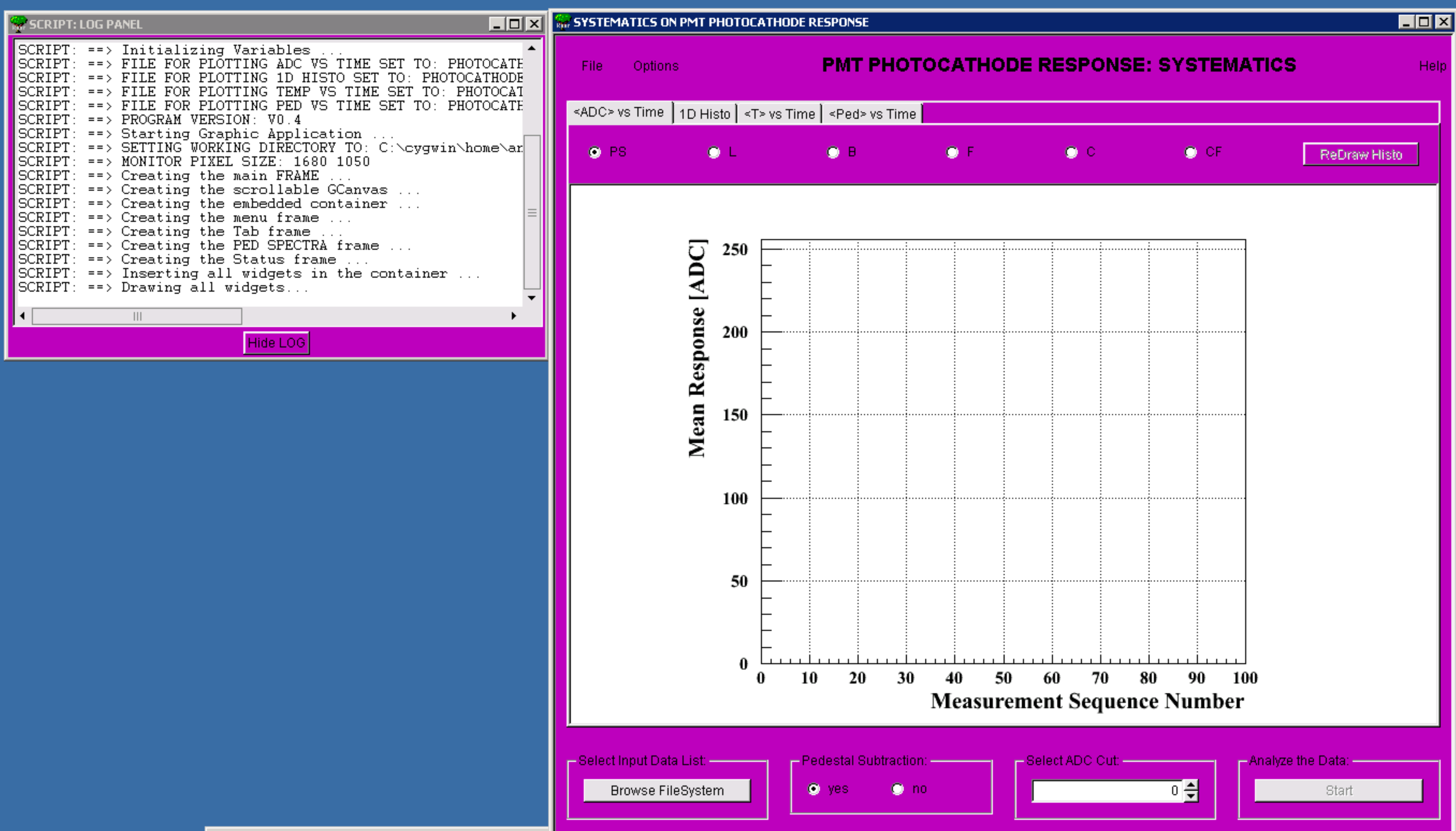}

   \vspace{-0.2cm}
   \caption{Startup of the PMT\_SYSTEMATICS GUI, along with 
            its LOG panel.}
   \label{fig:SYSTEMATIC_STARTUP}
\end{figure}

The program PMT\_SYSTEMATICS is a GUI with a layout similar 
to the PMT\_PhotoCathodeScan program, above described, 
and at startup shows up as in 
Fig.~\ref{fig:SYSTEMATIC_STARTUP}, 
with an additional small canvas dedicated to monitor the 
LOG activities.
To help the user in easily distinguish the two analysis programs 
(when both are running), a different color (violet) was chosen 
for the GUI dedicated to systematic studies. 

To start the analysis the user should select the input data list
via the button \underline{\it Select Input Data List} in the 
bottom left side of the canvas; the selection allows only to 
choose  one  *.list file, which contains a list of the *.cvs files
generated by the DetScan program. This list should be prepared in 
advance by the analyzer, and should contain in each line the 
correct path of a *.cvs file. That location (directory) 
should also contain all the files generated by the DetScan
needed, as for PMT\_PhotoCathodeScan program, to retrieve 
all necessary information for the analysis.

If an error in finding the list file or one of the files in 
the list is encountered, then that error will be dumped in the LOG 
panel. Instead, if no error appears  the button 
\underline{\it Start} will get 
activated and the data can be analyzed by pressing it.

Also here, after selecting the data file to process the user has the option
whether to subtract the pedestal calculated by the DetScan (in the
future it could be calculated directly from its spectrum) and to select a
lower-level ADC cut. The analysis is performed 
with the selected options. In case, a different set of options is
needed, then the analysis should be restarted.

The plots showed in the tabs can be saved in external files. This
possibility is  displayed in the menu option \underline{\it File}. 
The default graphic format is png but it can be changed via the menu 
option \underline{\it Options/Config Format for Pictures}.
In the same menu item \underline{\it Options} one can choose to show 
the legends in the plots, and to show the LOG file. At the end of the 
menu bar the option \underline{\it Help} gives the possibility to see 
the version history of the program and the reference to its author.

\begin{figure}[t!]
  \hspace{-1cm}
      \includegraphics[height=13cm, width=14.cm]
                      {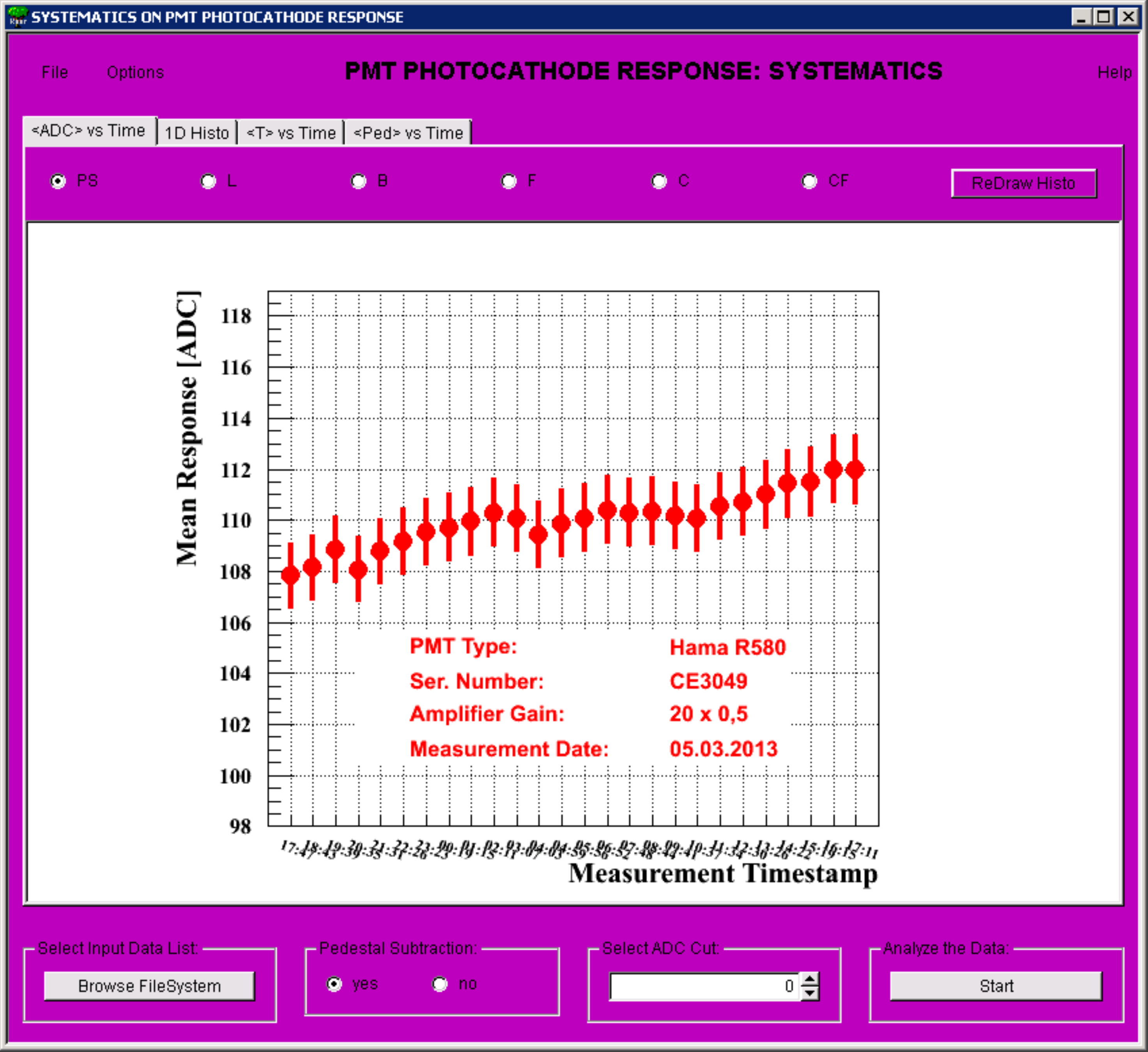}

  \vspace{-0.2cm}
  \caption{Behavior of the mean values of the temperature 
           measured inside and outside the scan box
           during the series of photocathode scans.
           for the Hamamatsu R580 (serial number CE3049).}
  \label{fig:SYSTEMATICS_MEANS}
\end{figure}

The Tab ({\underline {\it <ADC> vs Time}}) 
shows the mean amplitude of 
a scan for the series of measurements inserted in the selected list.
The program allows to change the visualization using different 
plotting options, as displayed in the radio-buttons
widgets in the top side of the tabs. In this case there is no need to 
restart the analysis, but it is sufficient to redraw the histogram with 
the button \underline{\it ReDraw Histo}. 
An example of a series of measurements covering one entire day 
(on March 2013) is shown in 
Fig.~\ref{fig:SYSTEMATICS_MEANS}. 

The measurements presented here and in the following were performed 
with the photomultiplier Hamamatsu R580 (serial number CE3049). 
A small trend during the investigated day of approximately $4\%$ is 
visible, although in most of the measurements the system appears quite 
stable. The statistical uncertainty of the mean is calculated out of 
each one-dimensional histogram containing the measured intensity 
for all scan-points as RMS/$\sqrt(N_{\textit{points of scan}}-1)$.
 
\begin{figure}[t!]
  \hspace{-1cm}
      \includegraphics[height=13cm, width=14.cm]
                      {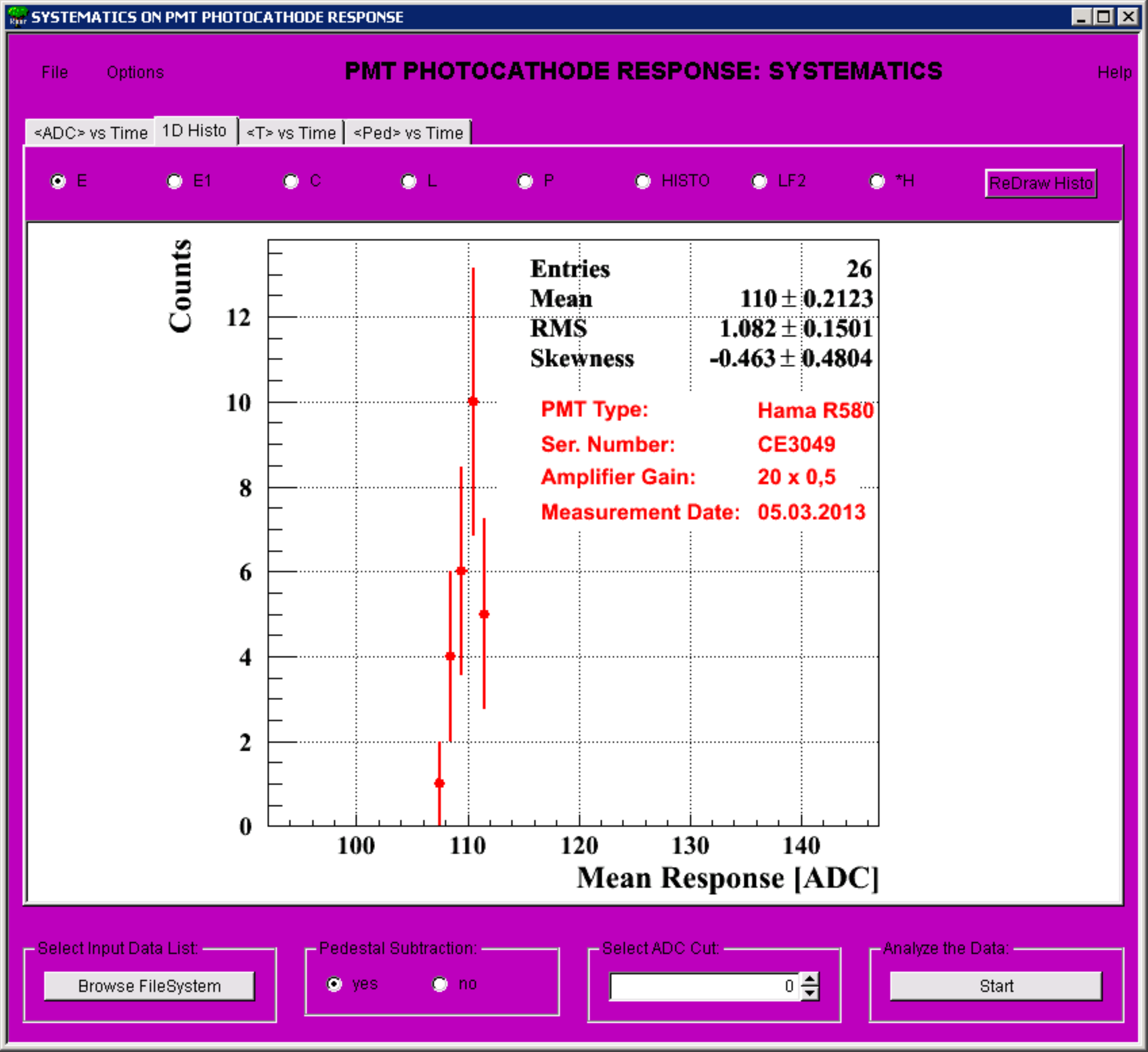}

  \vspace{-0.2cm}
  \caption{One-dimensional distribution of the mean values of the 
           intensity measured 
           during the series of the photocathode scans.}
  \label{fig:SYSTEMATICS_ONEDIM}
\end{figure}

The Tab {\underline {\it 1D Histo}} shows the distribution of all mean values 
shown in the Tab ({\underline {\it <ADC> vs Time}}).
For the above mentioned series of measurements the one-dimensional 
histogram is presented in Fig.~\ref{fig:SYSTEMATICS_ONEDIM}.
%

The Tab {\underline {\it <T> vs Time}} shows the mean temperature 
for the same photocathode scans. The temperature is measured
inside the scan light-tight box, and in the room containing the 
scanning device. The room is typically kept at a relatively stable 
temperature by a cooling system. For every scan, during each step 
of the LED on the photocathode surface the temperature is measured, 
and the statistical mean of all these measurements characterizes 
here the presented mean value. 

\begin{figure}[t!]
  \hspace{-1cm}
      \includegraphics[height=13cm, width=14.cm]
                      {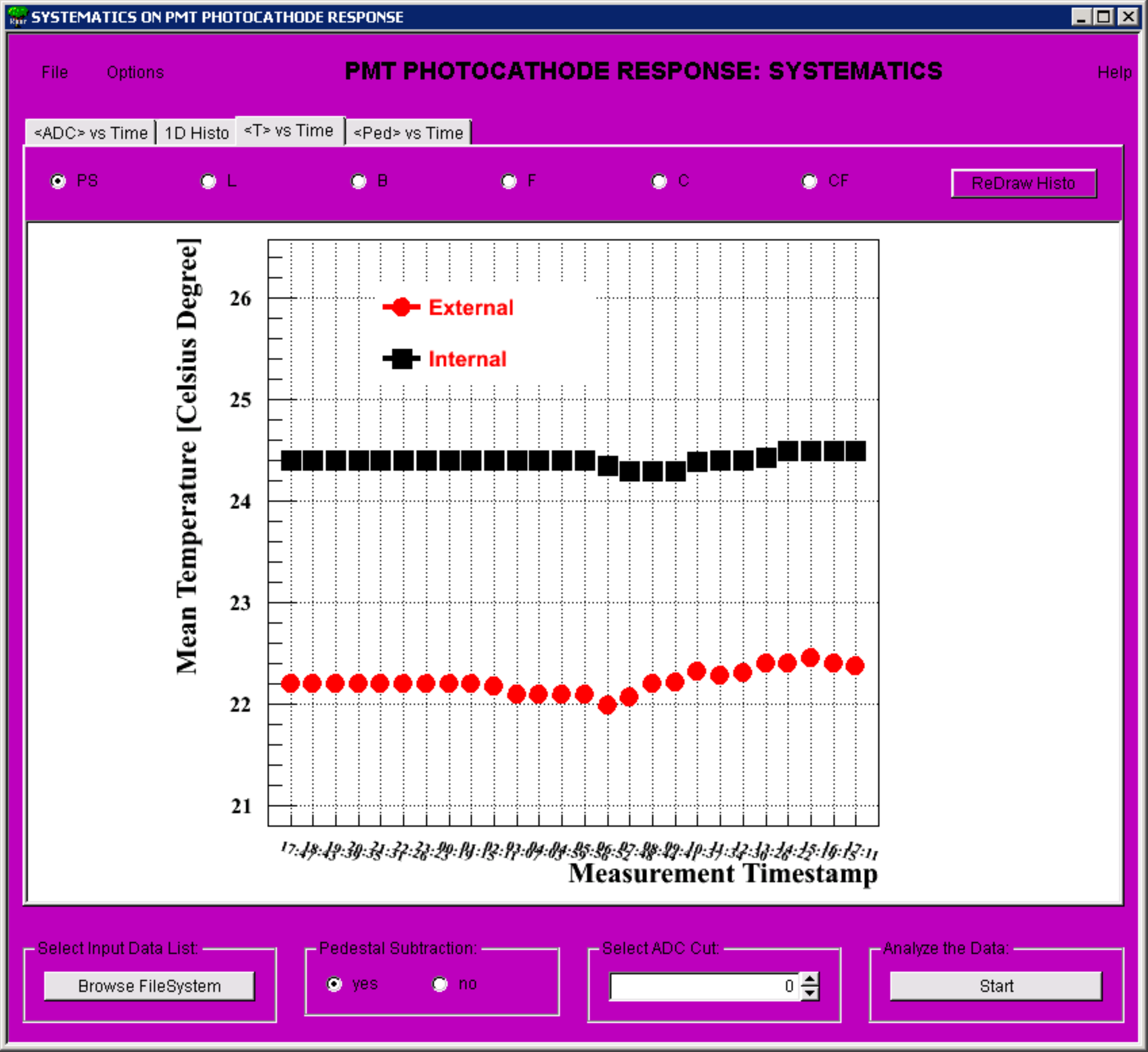}

  \vspace{-0.2cm}
  \caption{Behavior of the mean values of the temperature 
           measured inside and outside the scan box
           during the series of photocathode scans.}
  \label{fig:SYSTEMATICS_TEMP}
\end{figure}

For the considered series of measurements the behavior of 
the "inner" and "outer" temperature 
is shown in Fig.~\ref{fig:SYSTEMATICS_TEMP}. 
A small change of the 
temperature, in the order of fraction of one degree Celsius, is observed.
This tiny change can hardly be considered the source of the 
observed relative $4\%$ maximum deviation found in the response means. 

\begin{figure}[t!]
  \hspace{-1cm}
      \includegraphics[height=13cm, width=14.cm]
                      {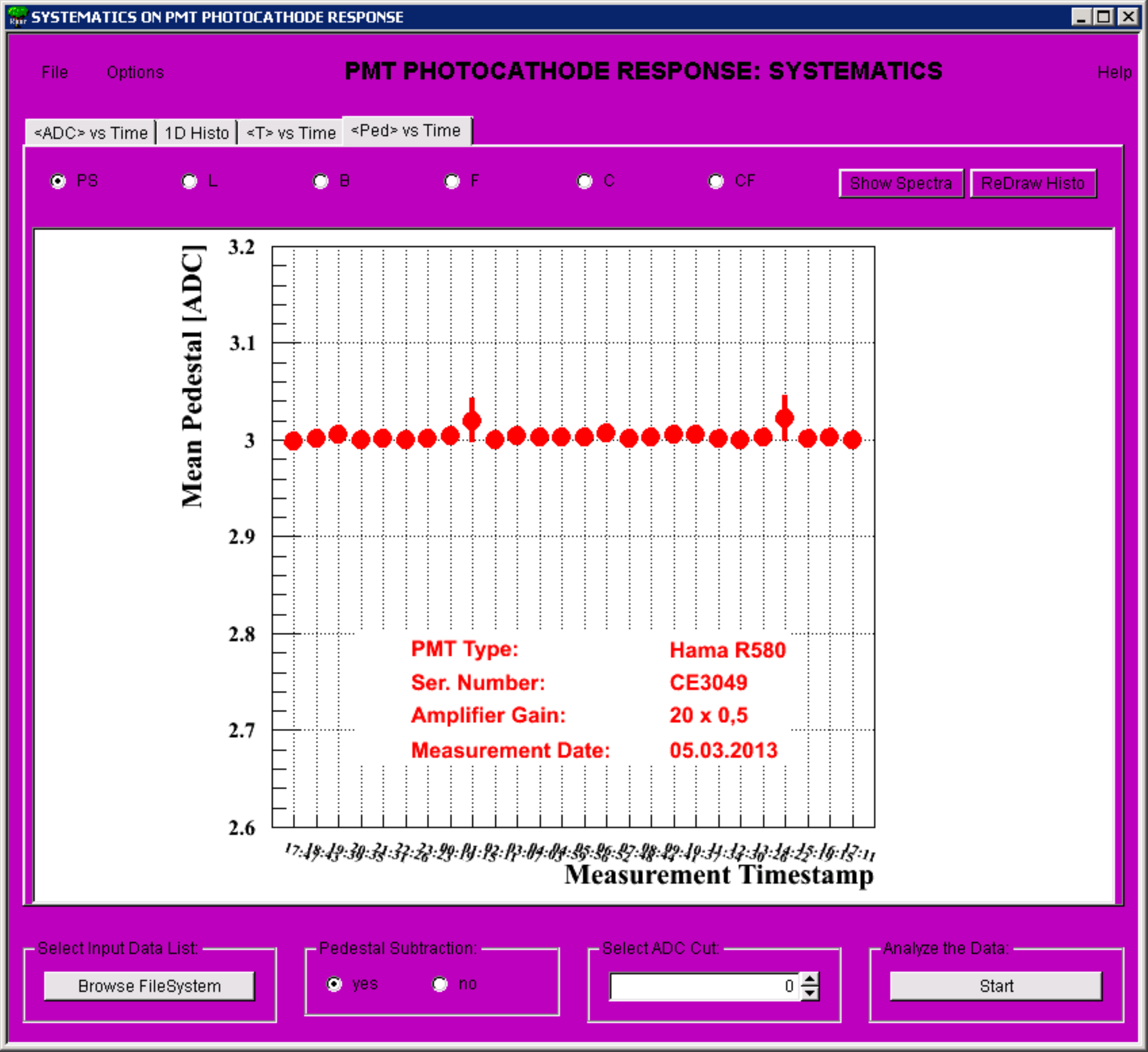}

  \vspace{-0.2cm}
  \caption{Distribution of the mean values of the pedestal measured 
           during the series of photocathode scans.}
  \label{fig:SYSTEMATICS_PED}
\end{figure}

At the beginning of each photocathode scan the LED is moved by the 
stepper motors far away from the PMT front-surface. This way, the 
pedestal of the system is obtained, measuring the signal in the 
PMT, and can be thus subtracted to the 
raw data in the offline analysis. 
Due to the system noise the distribution of the pedestal 
values (typically $1000$ LED pulses are sent) is not a delta function
but should typically follow a Gaussian distribution,  
and a mean value is thus calculated, 

The Tab {\underline {\it <Ped> vs Time}} shows the behavior the 
mean pedestal during the measurements. For the series of measurements
here considered as example, Fig.~\ref{fig:SYSTEMATICS_PED} shows a very 
good stability of the pedestal. 
%

\section{Conclusions and Outlook}
%
We have presented in this note new software tools for the 
analysis of the photocathode response of vacuum photomultipliers
(using the data accumulated by the DetScan program already available
at ZEA-2).
These programs performs a deeper investigation of the 
PMT data quantifying the level of homogeneity response, 
with and without the application of a lower level cut
and of the pedestal subtraction (when the pedestal has been 
measured). In addition, in case of multi-anode PMTs the 
gain mismatch of the individual channels can be investigated,
allowing to bring the non-homogeneity within few percents,

This code was already used to investigate the relative gain of
several Hamamatsu vacuum photomultipliers R580 during the 
commissioning of the JUDIDT electronics for an Anger Camera 
prototype~\cite{JUDIDT}.

The program to monitor the stability of the measurement
showed a systematical uncertainty of some percents, whose 
source has not been spotted yet. The blue-light LED
system could be a possible source of the observed "long-term" 
instability. 

In principle, the long-term behavior of additional variables,
e.g., the system noise, and two-dimensional correlation plots 
could be added in dedicated tabs, enlarging the control of the 
measuring system.

A natural extention of this software suite (not possible here
due to time constraints) would be the 
inclusion of an extra GUI dedicated to compare the measurements 
of different PMTs to investigate the relative gain.

%
\begin{quotation}
  \begin{center}
      {\bf Acknowledgments}
      \vspace{0.5cm}
  \end{center}
 The author gratefully acknowledges 
 U.\ Clemens, R.\ Engels and C.\ Wesolek
 for their valuable technical contribution and 
 suggestions to the work here presented. 

\end{quotation}



\end{document}